# Highly Complex Magnetic Structures Resulting From Hierarchical Phase Separation in AlCo(Cr)FeNi High Entropy Alloys

Qianqian Lan[1,2,*], András Kovács[1], Jan Caron[1], Hongchu Du[1,3], Dongsheng Song[1], Sriswaroop Dasari[4], Bharat Gwalani[4], Varun Chaudhary[5], Raju V. Ramanujan[5], Rajarshi Banerjee[4,5] and Rafal E. Dunin-Borkowski[1]

[1]*Ernst Ruska-Centre for Microscopy and Spectroscopy with Electrons and Peter Grünberg Institute, Forschungszentrum Jülich, 52425 Jülich, Germany*
[2]*School of Materials Science and Engineering, Tsinghua University, Beijing, 100086, China*
[3]*Central Facility for Electron Microscopy, RWTH Aachen University, 52074 Aachen, Germany*
[4]*Department of Materials Science and Engineering, University of North Texas, Denton, TX 76201, USA*
[5]*School of Materials Science and Engineering, Nanyang Technological University, Singapore 639798, Singapore*

## Abstract

Magnetic high entropy alloys (HEAs) are a new category of high-performance magnetic materials, with multi-component concentrated compositions and complex multi-phase structures. Although there have been numerous reports of their interesting magnetic properties, there is very limited understanding about the interplay between their hierarchical multi-phase structures and their local magnetic structures. By employing high spatial resolution correlative magnetic, structural and chemical studies, we reveal the influence of a hierarchically decomposed B2 + A2 structure in an AlCo$_{0.5}$Cr$_{0.5}$FeNi HEA on the formation of magnetic vortex states within individual A2 (disordered BCC) precipitates, which are distributed in an ordered B2 matrix that is weakly ferromagnetic. Non-magnetic or weakly ferromagnetic B2 precipitates in large magnetic domains of the A2 phase, and strongly magnetic Fe-Co-rich interphase A2 regions, are also observed. These results provide important insight into the origin of coercivity in this HEA, which can be attributed to a complex magnetization process that includes the successive reversal of magnetic vortices.

**Introduction**

Magnetic materials are essential in a wide and growing variety of industrial, commercial, and residential applications, including rotating electrical machines, electric vehicles, wind turbines, transformers, power convertors, electronic article surveillance systems, magnetically coupled devices, sensors, and high frequency electromagnetic devices [1, 2]. To illustrate the importance of magnetic materials, we note that the energy utilization of electrical machines is a significant fraction of the total energy consumption in the world [3]. Even a 1 % increase in their efficiency would have a great impact on energy efficiency and on the reduction of $CO_2$ emissions. Hence, there is considerable interest in the development of improved magnetic materials, such as magnetic high entropy alloys (HEA), for next generation magnets.

HEAs are formed from near-equiatomic proportions of five or more elements [4, 5]. They are attracting significant attention because their rich compositional variety and phase space provides opportunities for discovering alloys that have outstanding mechanical and functional properties [6, 7]. HEAs can form single phases, comprising random solid solutions of their constituent elements. However, local variations in chemical composition and short-range order have a strong effect on dislocation movement and planar defect energy, leading to increased yield strength without compromising strain hardening and tensile ductility [8], just as for numerous multiphase alloys and composites in which increased heterogeneity results in improvements in properties [9]. Considering the huge number of possible combinations of elements and processing conditions that can lead to the formation of secondary phases, the possibilities for developing multi-phase systems with different functionalities is enormous [10].

The formation of compositional heterogeneities and secondary phases can also alter the magnetic properties of HEAa. In general, ferromagnetic HEAs that have face-centered cubic (FCC) and body-centered cubic (BCC) structures are based on Fe, Co and Ni. An exception is the hexagonal-close-packed (HCP) HoDyYGdTb HEA, which exhibits a rich and complex magnetic phase diagram below room temperature [11]. Although the mechanical properties of many HEAs are well studied and can be superior to those of conventional alloys, much less work has been performed to obtain an in-depth understanding of their magnetic properties.

Ferromagnetic HEAs have good mechanical properties and a wide range of tunable magnetic properties, ranging from soft to semihard to hard [12-14]. The magnetic characteristics of HEAs are highly sensitive to the compositions and morphologies of the constituent phases over various length scales [4, 15-17]. For example, the addition of paramagnetic or antiferromagnetic elements can induce phase segregation and decrease saturation magnetization $M_S$ [15], The addition of 25% Cr to FeCoNi can make the resulting FeCoNiCr alloy paramagnetic [18], while the introduction of Mn to FeCoNiCr can eliminate the energy difference between FCC and HCP structures, resulting in magnetic frustration [19].

An in-depth understanding of structure-magnetism correlations in HEAs has been limited by a lack of appropriate methods to characterize their phases and properties at the nanoscale and beyond. Most magnetic property studies of HEAs are based on conventional magnetization hysteresis loop (M-H) measurement of bulk forms of the HEAs, and are unable to reveal the local magnetic state and interactions between the constituent phases. The magnetic properties of highly heterogeneous alloys, in which each phase displays chemical and topological disorder, cannot be described as a composition-weighted average of the magnetic properties of the constituent phases [16, 17]. Specifically, the origin of the dramatic difference between the magnetic properties of AlCoFeNi and $AlCo_{0.5}Cr_{0.5}FeNi$ HEA is unexplored. An experimental understanding of the relationship between the local structure and magnetic texture of AlCoFeNi and $AlCo_{0.5}Cr_{0.5}FeNi$ HEAs is crucial to control their properties.

Here, we use three-dimensional (3D) atom probe tomography (APT) and transmission electron microscopy (TEM) methods, including the Fresnel mode of Lorentz TEM and off-axis electron holography (EH), to investigate the influence of complex hierarchical phase separation of the A2 phase (disordered BCC) and the CsCl-structured B2 phase (ordered BCC) on local variations in the magnetic structure of AlCo(Cr)FeNi HEAs. We study the local magnetic characteristics of the individual phases, including their saturation magnetic induction and coercivity. Our results provide direct experimental measurements of correlations between structure and magnetic properties in HEAs with nm spatial resolution. This information is essential for understanding complex magnetism in multi-phase AlCo(Cr)FeNi alloys, as well as for the design of new HEAs with unique tailored magnetic properties.

**Results**

We first report the magnetic properties of AlCoFeNi and AlCo(Cr)FeNi bulk HEAs, followed by TEM studies of the structure and magnetism of AlCoFeNi. The multi-length-scale structure of AlCo(Cr)FeNi is then investigated. Two regions (R1, R2) were identified, within which a heterogeneous phase distribution was observed. The static and dynamic magnetic behavior of R1 and R2, as well as the phases present in these regions, was studied in detail.

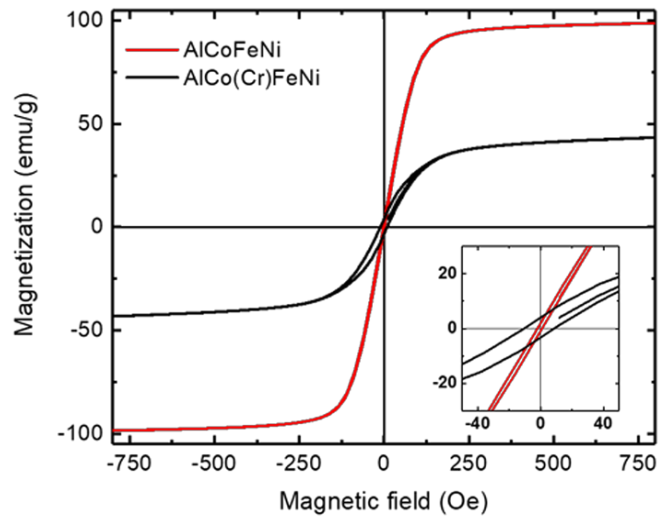

**Figure 1** Magnetization (*M*) *vs* applied magnetic field (*H*) measured at a temperature of 300 K for AlCoFeNi and AlCo(Cr)FeNi HEAs. The inset shows a magnified view of the central part of the hysteresis loop.

1. **Magnetic properties of heat-treated AlCoFeNi and AlCo(Cr)FeNi bulk HEAs.**

Figure 1 shows magnetization *vs* magnetic field (*M-H*) hysteresis loops measured at 300 K of heat-treated AlCoFeNi and AlCo(Cr)FeNi alloys. The loops suggest soft ferromagnetic properties. AlCoFeNi HEA has a saturation magnetization ($M_S$) of 99.8 emu/g and a coercivity ($H_c$) of 1.45 mT. When half of the Co is replaced by Cr in the AlCo(Cr)FeNi HEA, $M_S$ decreases to 46.2 emu/g, while $H_c$ increases to 9.56 mT. The HEAs are almost fully magnetically saturated at fields of 500 mT. Cr substitution and annealing induce complex hierarchical phase morphology in heat-treated AlCo(Cr)FeNi [13, 20], defining the resulting magnetic properties. It is therefore important to understand the influence of each phase on the magnetic properties of AlCoFeNi and AlCo(Cr)FeNi HEAs.

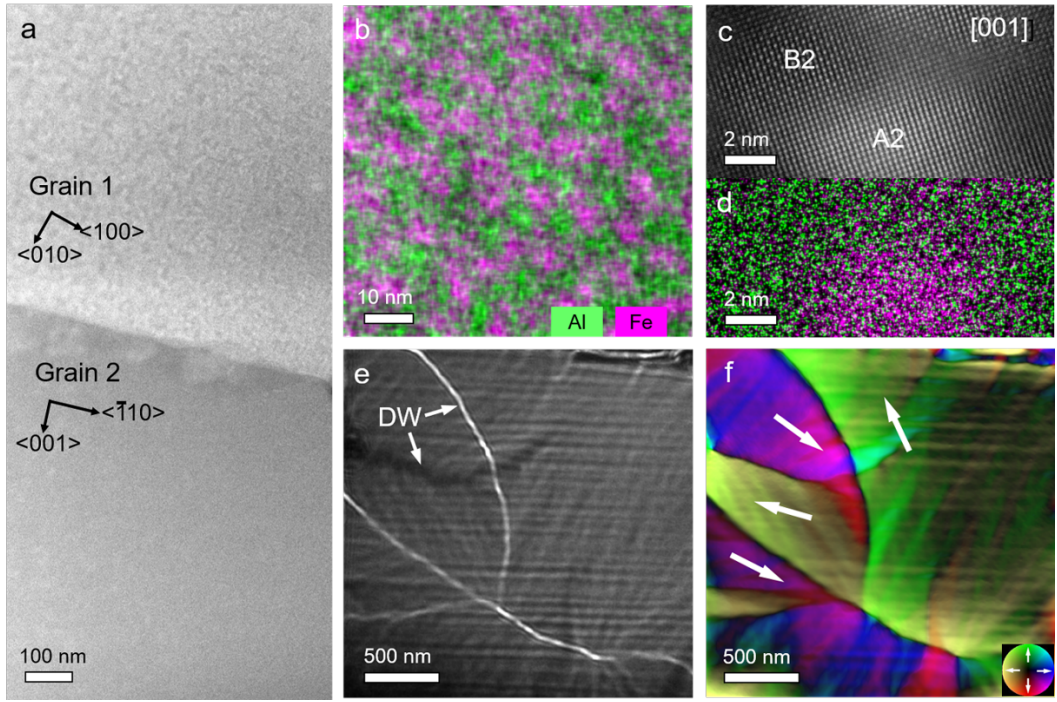

**Figure 2 Structural and magnetic texture of the AlCoFeNi HEA.** (a) HAADF STEM image of two grains, with grain 1 aligned to a zone axis. The orientations of the grains are indicated. (b) Combined Al+Fe elemental map obtained from grain 1 using STEM EDXS imaging. (c) Atomic-resolution HAADF STEM image and (d) corresponding EDX map. (e) Overfocus (100 $\mu$m) Fresnel defocus image recorded from grain 2 and (f) corresponding magnetic induction map reconstructed using the transport-of-intensity equation. The color wheel indicates the direction and magnitude of the projected in-plane magnetic induction.

## 2.    Microstructural and magnetic properties of heat-treated AlCoFeNi.

The AlCoFeNi HEA has a polycrystalline structure with a grain size of above 100 $\mu$m. Figure 2a shows a high-angle annular dark-field (HAADF) scanning TEM (STEM) image of a grain boundary region between two adjacent AlCoFeNi grains. Grain 1 was aligned in the electron microscope to the closest crystallographic zone axis, at which an (001) direction was parallel to the incident electron beam direction. In this projection, chemically sensitive contrast reveals inhomogeneities. 3D APT studies of AlCoFeNi have previously revealed the presence of nm-sized Fe-Co-rich A2 precipitates in an Al-Ni-rich B2 matrix [13].

Figure 2b shows Al and Fe elemental distributions measured using energy-dispersive X-ray spectroscopy (EDXS) in grain 1, confirming the presence of compositional variations. The average sizes of the Al-rich and Fe-rich regions are approximately 10 nm. Figures 2c and 2d show an atomic-resolution image and a corresponding elemental map of the Al-rich B2 phase and the Fe-rich A2 phase. Chemical ordering in the B2 phase is visible in the HAADF STEM image. The structures are mostly coherent, with occasional dislocations at the interface due to the lattice misfit.

The magnetic structure at remanence was initially studied using Fresnel defocus imaging in Lorentz mode in magnetic-field-free conditions. Figure 2e shows a representative overfocus Fresnel image recorded from grain 2. This region contains large magnetic domains, which are separated by magnetic domain walls that appear as black and white lines in the image. The ripple-like contrast variations in the domains originate from small-angle magnetization variations. Figure 2f shows an approximate representation of the projected in-plane magnetic induction obtained from a pair of such defocused images using the transport-of-intensity equation [21, 22]. The magnetic domain configuration in the thin specimen was observed to rearrange upon applying a magnetic field as small as 5 mT perpendicular to the specimen using the conventional electron microscope objective lens. This observation suggests isotropic soft ferromagnetic behavior, in which the phase separated structure is not strong enough to pin magnetic domain walls.

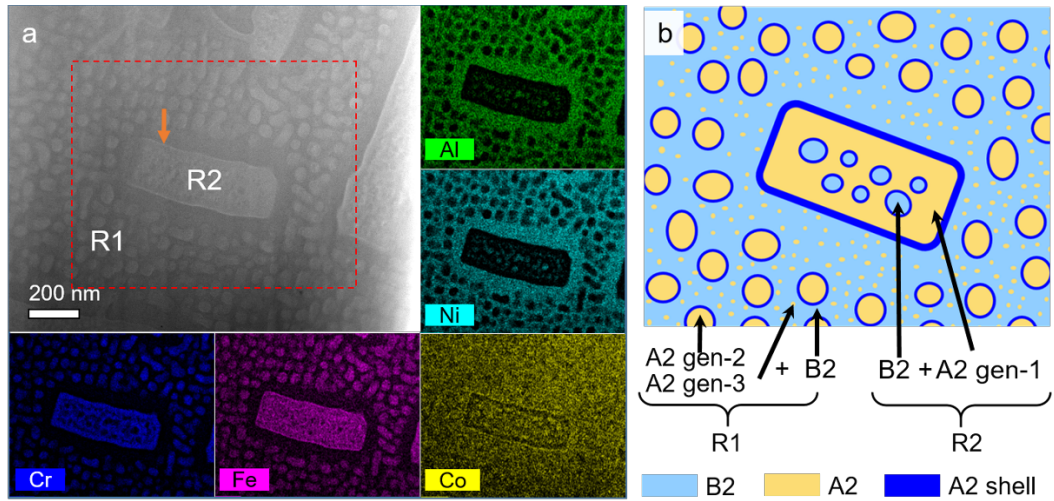

**Figure 3 Phase separation in the AlCo(Cr)FeNi HEA.** (a) HAADF STEM image of the multi-scale hierarchical B2 and A2 structure, containing characteristic R1 and R2 regions. EDXS elemental maps of Al, Cr, Fe, Co and Ni recorded from the marked rectangular area are shown around the main frame. An orange arrow marks an A2 shell around an R2 island. (b) Schematic diagram of the R1 and R2 phase arrangements. In the R1 phase, fine A2 gen-3 and medium A2 gen-2 precipitates form in a B2 matrix. The A2 gen-2 precipitates are covered by an A2 shell. In region R2, B2 precipitates form in an A2 gen-1 matrix covered by an A2 shell.

### 3. Multi-length-scale structure of AlCo(Cr)FeNi.

Figure 3a shows that the annealed AlCo(Cr)FeNi HEA specimen has a strikingly different structure from that of the AlCoFeNi alloy, comprising an Al-Ni-Co-rich B2 phase, Fe-Cr-Co-rich A2 regions and their combinations. The different length scales of the A2 phases are referred to here as a) coarse A2 gen-1 (with an average size of 1-5 $\mu$m), b) medium-scale A2 gen-2 (50-150 nm) and c) fine-scale A2 gen-3 (<10 nm). Two characteristic regions are labeled R1 (A2 gen-2 + A2 gen-3 + B2 matrix) and R2 (A2 gen-1 + B2 matrix). An additional A2 phase (A2 shell) is observed between the B2 and A2 phases (see below), taking the form of a few-nm-thick shell around the A2 gen-1 and A2 gen-2 phases. Figure 3b shows a schematic diagram of the constituent phases and regions in the AlCo(Cr)FeNi HEA specimen.

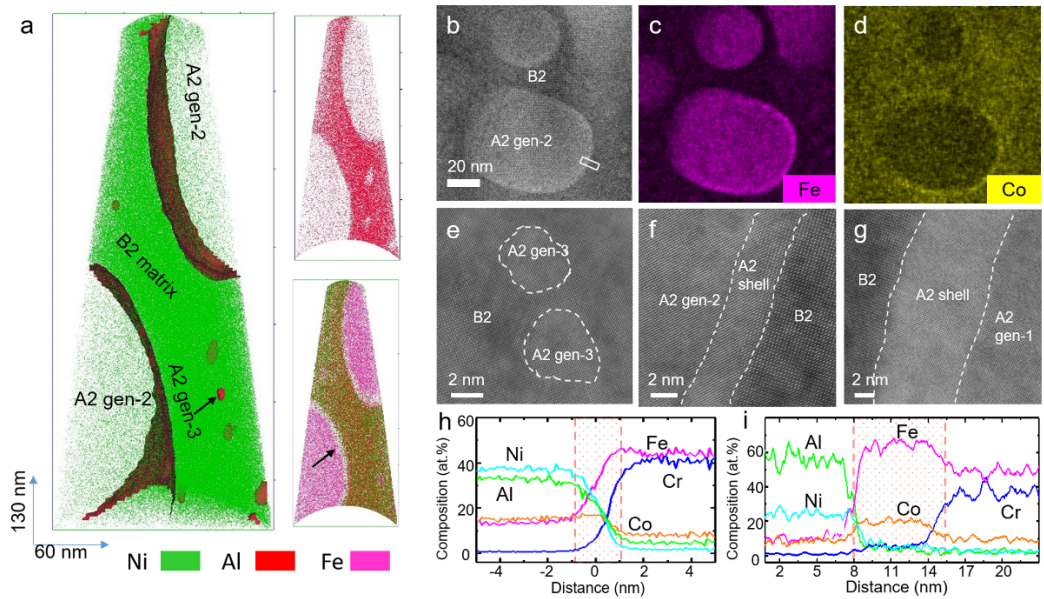

**Figure 4 Microstructure and chemical composition of the annealed AlCo(Cr)FeNi HEA.** (a) 3D APT reconstruction of A2 gen-2 and gen-3 precipitates in a B2 matrix in region R1. An Fe-Co-rich shell forms around the A2 gen-2 precipitates. The colors represent Al (red), Ni (green) and Fe (magenta). (b-d) HAADF STEM image and corresponding Fe and Co elemental maps measured using EDXS imaging. Fe and Co enrichment is evident in the interface region of the two phases. (e-g) Atomic-resolution HAADF STEM images of A2 gen-3 in the B2 matrix, the A2 shell around the A2 gen-2 precipitates (marked by a white rectangle in Fig. 4b) and the A2 shell around an A2 gen-1 island (marked by an orange arrow in Fig. 3a), respectively. (h, i) Compositional profiles across the A2 shells around the A2 gen-2 phase and the A2 gen-3 phase, respectively. Fe and Co enrichment is observed in the shells.

Figure 4 shows TEM and APT measurements of the shell A2 phase. Figure 4a shows an APT reconstruction of the elemental distribution in a needle-shaped specimen of region R1 in the AlCo(Cr)FeNi HEA, revealing small Fe-Cr-rich A2 gen-3 precipitates and large A2 gen-2 precipitates surrounded by Fe-Co-rich A2 shells in a B2 matrix. A detailed description of the composition of the shell is presented in the Supplementary Information. Figures 4b-4d show an HAADF STEM image and corresponding Co and Fe EDXS elemental maps of the shell region. The enhancement in Co and Fe at the edges of the A2 gen-2 precipitate appears to be discontinuous, perhaps because of the geometry of the thin TEM specimen, from which part of the precipitate may have been removed by ion milling. Figure 4e shows an atomic-resolution HAADF STEM image of A2 gen-3 precipitates in the B2 matrix. Figures 4f and 4g show atomic-resolution HAADF STEM images of the A2 structure of the shell, whose thickness is 2 nm around the A2 gen-2 precipitates and 7.5 nm around the A2 gen-1 island. The measurements are consistent between the APT result and the EDXS images shown in Figs 4h and 4i. The A2 shell is rich in Fe and Co, whereas little or no Al or Cr is present inside it. Atomic-resolution HAADF STEM images of the interface region are discussed in detail in the Supplementary Information.

## 4. Magnetic structure of AlCo(Cr)FeNi (I) − A2 gen-2 precipitates in the B2 matrix (region R1).

### 4.1 Static magnetic structure

Highly heterogeneous systems, in which each region contains both chemical and topological disorder, such as heat-treated AlCo(Cr)FeNi HEAs, typically have complex magnetic properties in the constituent phases that cannot be understood from bulk magnetic measurements alone. We studied the local magnetic properties of the A2 and B2 phases in the R1 and R2 regions of the AlCo(Cr)FeNi specimen using Fresnel defocus imaging and off-axis EH. The latter technique provides a direct quantitative measurement of the phase shift of the electron wave that interacted with the specimen, from which the local magnetic state of the region of interest can be determined with nm spatial resolution [23]. The total electron optical phase shift contains electrostatic and magnetic contributions that need to be separated to measure the magnetic field distribution in the specimen. In the absence of electron-beam-induced charging, the electrostatic phase shift is proportional to the projected mean inner potential (MIP) of the specimen, which depends on its composition, density and ionicity.

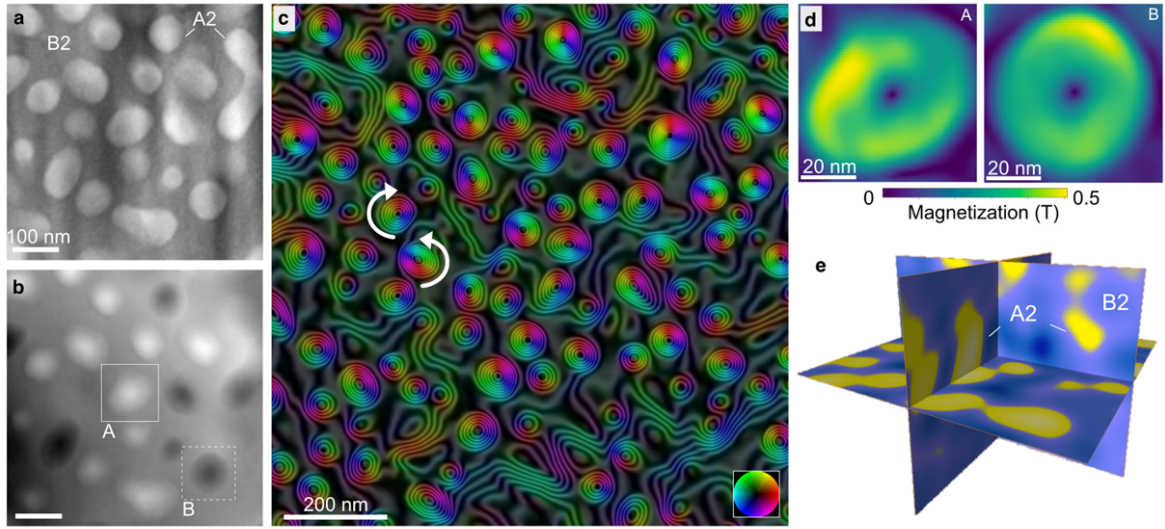

**Figure 5 Magnetic structure of region R1 containing A2 precipitates in a B2 matrix.** (a) Mean inner potential and (b) magnetic contribution to the phase shift ($\phi_M$) recorded using off-axis EH with the specimen in a magnetic remanent state. The magnetic phase shift ($\phi_M$) in the A2 gen-2 precipitates is either bright or dark, as a result of the presence of a magnetic vortex in each precipitate. The marked A2 gen-2 precipitates (A and B) are further analysed in (d). (c) Large-field-of-view magnetic induction map derived from the magnetic contribution to the phase shift, showing clockwise and counterclockwise magnetic vortex states in the A2 precipitates. The phase contour spacing is $2\pi/24$ rad. (d) Projected in-plane magnetization ($M_{xy}$) in the A2 precipitates marked in (b) determined from the magnetic contribution to the phase shift using model-based iterative reconstruction. An upper limit for the diameter of the magnetic vortex core, which has an out-of-plane magnetic field orientation, is 8 nm. (e) Sections showing embedding of A2 precipitates in the B2 matrix extracted from a tomographic reconstruction obtained from a tilt series of ADF STEM images.

Figure 5a shows the MIP contribution to the electron optical phase shift in region R1 measured using off-axis EH at magnetic remanence. In this image, the A2 gen-2 precipitates, which have close-to-spherical morphologies and diameters of between 50 and 120 nm, appear brighter than the surrounding B2 matrix, as they have a higher mean atomic number per unit volume. Figure 5b shows the corresponding magnetic contribution to the phase shift $\phi_M$, which provides a measure of the in-plane component of the magnetic induction within and outside the specimen integrated in the electron beam direction[24]. Bright or dark contrast is visible within the boundaries of the A2 gen-2 precipitates, which are each surrounded by a thin Fe-Co-rich A2 shell. Figure 5c shows a magnetic induction map obtained by generating contour lines and colors from the magnetic contribution to the phase shift and its gradient, respectively. This image reveals that each A2 precipitate contains a magnetic vortex, with the magnetic field rotating either clockwise or counterclockwise. Similar 3D magnetic vortex states have been observed in sub-100-nm spherical Fe-Ni particles without strong magnetocrystalline anisotropy[25, 26]. In region R1, the A2 precipitates are well separated by the B2 matrix, preventing dipolar interactions between individual crystals. The ratio of clockwise to counterclockwise magnetic vortices is approximately 1:1 at remanence after saturating the specimen using the microscope objective lens, suggesting energetically-independent magnetic states that are only weakly coupled to the surrounding phases. The vortices are not associated with significant stray magnetic fields that could be measured at remanence using either bulk magnetometry or surface-sensitive magnetic characterization techniques.

A model-based iterative reconstruction algorithm[27] was used to convert the measured magnetic phase images $\phi_M$ into maps of projected in-plane magnetization $M_{xy}$, as shown in Fig. 5d for the two A2 gen-2 precipitates marked in Fig. 5b. The magnetization direction of the vortex core is parallel to the incident electron beam direction in the center of each precipitate and does not contribute to the projected in-plane magnetization map in this region. In general, each magnetic vortex core can point either up or down magnetically, irrespective of the vortex rotation direction[28]. The magnetization direction of the vortex core cannot be detected from these images, as EH is sensitive to the component of the magnetic field that is perpendicular to the incident electron beam direction. An upper limit for the magnetic vortex core diameter was measured to be ~8 nm by fitting a Gaussian function to the projected in-plane magnetization distribution. On the assumption that the specimen is magnetically active through its entire thickness, the magnitude of the in-plane magnetic induction (Fig. 5d) peaks at 0.5±0.1 T in the A2 gen-2 precipitates.

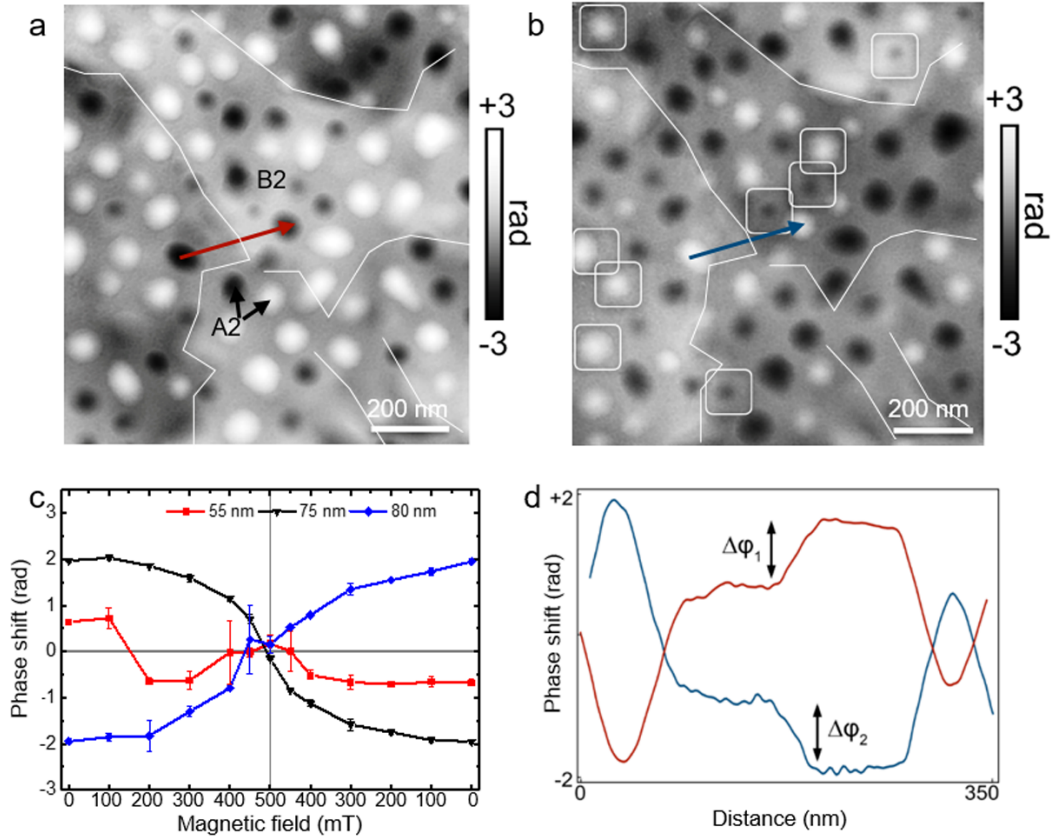

**Figure 6**    **Magnetic switching of A2 gen-2 precipitates in the R1 phase in the AlCo(Cr)FeNi HEA.** (a, b) Magnetic contribution to the phase shift $\phi_m$ recorded after returning to remanence from opposite out-of-plane fields of -500 and 500 mT, respectively. The localized regions of dark and bright contrast correspond to clockwise or counterclockwise magnetization rotation directions in individual A2 precipitates. Contrast reversal of the A2 gen-2 precipitates is associated with a change in the magnetic field rotation direction of the vortices. In (b), marked precipitates (white squares) retained their magnetization rotation direction from that observed in (a). (c) Magnitude and sign of the local change in magnetic phase shift ($\phi_M$) of individual A2 gen-2 precipitates of different size plotted as a function of out-of-plane magnetic field. Positive values are associated with counterclockwise magnetization rotation directions. (d) Line profiles of magnetic phase shift extracted from (a, b) between two A2 gen-2 precipitates, as indicated by red and blue arrows. The step in the phase shift $\Delta\phi$ indicates a projected in-plane magnetic field contribution from the B2 matrix.

The Fe-Cr-Co-rich A2 gen-2 precipitates are covered by Fe-Co-rich A2 shells (Fig. 4). The measured magnetic signal is therefore a superposition of contributions from the two A2 phases. Non-uniform magnetization distributions in some of the A2 precipitates may result from their "incomplete" morphologies in an ion-milled TEM specimen. 3D tomographic reconstruction was used to clarify the shapes and distributions of the A2 gen-2 precipitates in the B2 matrix from a tilt series of ADF STEM images[29]. Figure 5e shows sections through a tomographic reconstruction of region R1 that contains approximately spherical A2 gen-2 precipitates (yellow) in a B2 matrix (blue). Some of the precipitates intersect the TEM specimen surface and are incomplete.

## 4.2 *In-situ* magnetic switching of BCC gen-2 precipitates in the B2 matrix (region R1).

The magnetic switching properties of A2 gen-2 precipitates in the B2 matrix (region R1) were studied by applying magnetic fields perpendicular to the specimen plane. *In situ* magnetization reversal was performed by applying magnetic fields of up to 1.5 T using the conventional microscope objective lens. Figures 6a and 6b show magnetic phase images of region R1 recorded after returning to remanence from opposite out-of-plane fields of -500 and 500 mT, respectively. The proportion of clockwise and counterclockwise magnetic vortices in the A2 gen-2 precipitates in region R1 was measured from the magnetic phase images. A comparison of Figs 6a and 6b shows that the majority of the A2 gen-2 precipitates changed their magnetic field rotation direction. Within the field of view of approximately 1.4 $\mu$m × 1.4 $\mu$m, 46 counterclockwise and 34 clockwise vortices are visible in Fig. 6a, whereas 35 counterclockwise and 45 clockwise vortices are visible in Fig. 6b. White squares in Fig. 6b mark precipitates that retained the same sense of magnetic rotation as in the initial remanent state.

Figure 6c shows the magnitude of the local change in magnetic phase shift $\phi_m$ recorded from representative A2 gen-2 precipitates of different size, plotted as a function of applied magnetic field from 0→500→0 mT. The positive (or negative) sign of $\phi_m$ is related to a counterclockwise (or clockwise) magnetic field rotation direction in the vortices. Two large A2 precipitates with diameters of 75 and 80 nm show a continuous decrease in magnetic phase shift close to zero as the applied magnetic field is increased to 500 mT, suggesting that the internal field in the precipitates becomes aligned parallel to the electron beam direction. As the applied magnetic field is decreased (Fig. 6c), the magnetic phase shift recovers, but with opposite sign, indicating that the magnetic vortex now has a rotation sense opposite to the original rotation direction. Different switching behavior was observed for a small A2 precipitate with a diameter of 55 nm (Fig. 6c). The initial rotation sense is counterclockwise at 0 mT, changes sign at 200 mT and decreases gradually to zero as the applied magnetic field increases to 500 mT. A possible scenario is that the magnetic field direction of the vortex core was aligned antiparallel to the saturating magnetic field. At 200 mT, the vortex core switches to become aligned with the saturating field, which also changes the vortex rotation direction. As the applied magnetic field is increased further, this state becomes aligned with the applied field direction and the magnetic phase shift approaches zero. On decreasing the applied magnetic field, a magnetic vortex forms again at 400 mT and remains stable as the applied magnetic field is reduced to zero.

The magnetic nature of the B2 matrix in region R1, which contains more than 60% Fe, Co and Ni according to APT and EDXS measurements (Fig. 4), is now discussed. Figure 6d shows a plot of the magnetic phase shift $\phi_m$ in the B2 matrix in region R1 before and after magnetization reversal, revealing a region with a gradient in phase and an associated step $\Delta\phi_m \sim 1.3$ rad. The greatest intensity maxima and minima in the plot correspond to A2 gen-2 magnetic vortices that changed their rotation direction during switching. These A2 phases serve as a reference for studying the magnetic phase contrast change within the matrix region. Changes in the sign of the magnetic phase shift $\phi_m$ indicate that part of the Al-Ni-Co-rich B2 matrix is also magnetized in the plane of the specimen, is ferromagnetic and reverses in sign magnetically. Dipolar or magnetostatic interactions [30] between the A2 precipitates are expected to be affected by the nature of the surrounding magnetic phases and interphase boundaries. It is noteworthy that the same magnetic contrast is observed in the B2 matrix, but with a sign change in the magnetically-switched region R1 (Fig. 6).

The *in situ* magnetic switching experiments reveal details about the magnetic properties of region R1 in annealed AlCo(Cr)FeNi HEAs that contain A2-type precipitates in a B2 matrix. However, the lack of information about the core direction from off-axis EH experiments limits our understanding of the details of the process. The switching characteristics of the magnetic vortices depend on the sizes and shapes of the A2 gen-2 precipitates, the external magnetic field and coupling to the B2 matrix. Further analysis of this complex system requires comparisons of experimental measurements with atomistic spin dynamics or micromagnetic calculations that are beyond the scope of the present paper.

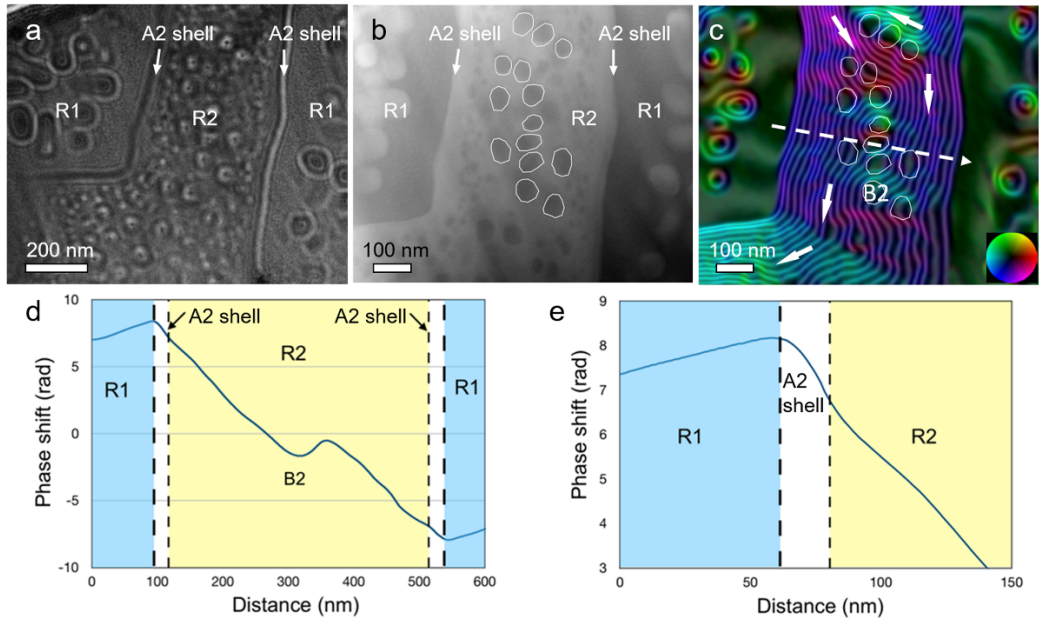

**Figure 7 Magnetic structure of a B2 + A2 solid solution mixture in regions R1 and R2 and the A2 shell of the AlCo(Cr)FeNi HEA.** (a) Fresnel defocus image recorded at remanence. The defocus value used was -200 $\mu$m. (b) Mean inner potential contribution to the phase measured using off-axis EH from the R2 (+A2 shell) region between two R1 regions. (c) Corresponding magnetic induction map. The phase contour spacing is $2\pi/16$ rad. Selected B2 inclusions in the A2 gen-1 matrix are marked with white frames in (b) and (c). The colors and arrows mark the projected in-plane magnetic field direction. (d) Line profile of the magnetic phase shift across an R2 region that includes A2 shell regions and a single B2 precipitate. The dip in the middle of the phase shift profile is associated with a weakly magnetic or non-magnetic B2 phase. (e) Line profile of the magnetic phase shift, in which the slope in an A2 shell region (0.075 rad/nm) is 25% higher than that in an R2 region (0.06 rad/nm).

**5. Magnetic properties of AlCo(Cr)FeNi (II) – B2 precipitates in the A2 gen-1 matrix (region R2).**

The AlCo(Cr)FeNi alloy contains regions (R2) of micrometer-sized Fe-Cr-Co-rich A2 gen-1 matrix with B2 precipitates. Microstructural and chemical studies show that an A2 shell is present between the B2 and A2 phases (Fig. 4). Figure 7a shows a Fresnel defocus image of an R2 island surrounded by R1 recorded at a defocus value of -200 $\mu$m. The magnetic nature of the R2 phase is apparent from the presence of dark and bright bands of contrast. The image also shows the limitation of Fresnel defocus imaging, as the phase boundaries give rise to strong fringing fields at their edges, making it difficult to interpret the magnetic state.

Figures 7b and 7c show the MIP contribution to the phase and a corresponding magnetic induction map recorded using off-axis EH. In Fig. 7b, the A2 gen-1 and gen-2 (bright contrast) and B2 (dark contrast) phases can be distinguished, as they have different mean atomic numbers per unit volume. Selected B2 precipitates are marked by white frames in both images. The color-coded magnetic induction map shows that the R2 region contains large magnetic domains. The magnetic field lines are either disrupted or missing at the B2 precipitates, suggesting that they are weakly magnetic or non-magnetic. The effect of the smaller (<50 nm) B2 precipitates on the magnetic field is less clear, as it can be masked by the signal from the A2 gen-1 matrix in the ~100-nm-thick TEM specimen.

Figure 7d shows a line profile of the magnetic phase shift across region R2, which contains a single B2 precipitate at its center. The weakly magnetic or non-magnetic Al-Ni-Co-rich B2 precipitate has a lower contribution to the magnetic phase shift and appears as a dip. Close inspection of the phase profile in Fig. 7e reveals a change in slope at the position of the A2 shell, suggesting a difference in its magnetic properties from those of the A2 gen-1 matrix. The slope of the phase in the A2 shell and the A2 gen-1 matrix in region R2 is 0.075 and 0.06 rad/nm, respectively, based on fitted linear functions. As the magnetic phase shift scales with the projected in-plane magnetic induction in the specimen, it can be inferred that the A2 shell has approximately 25% higher magnetization than the A2 gen-1 matrix in region R2. This difference is thought to result from the higher concentration of Cr in the A2 gen-1 matrix, which decreases the magnetization of the Cr-Fe-Co-rich A2 gen-1 matrix in the R2 region.

**Discussion**

Saturation magnetization and coercivity in AlCo(Cr)FeNi HEAs are affected by the presence of non-magnetic Al and antiferromagnetic Cr. The AlCoFeNi HEA contains nm-sized Fe-Co-rich A2 precipitates in an Al-Ni-rich B2 matrix. The saturation magnetic induction of AlCoFeNi of 0.987 T (99.8 emu/g) is associated primarily with ferromagnetic ordering of Fe, Co and Ni.

The effect of Cr substitution and heat treatment of AlCo(Cr)FeNi leads to decomposition of the alloy into an interesting duplex distribution of A2 and B2 phases in regions R1 and R2. Bulk measurements of structure and magnetic properties are not able to resolve and provide an understanding of the separate contributions of each phase. On the other hand, the high spatial resolution analyses presented here reveal the intimate relationship between their structure and magnetic properties. Upon annealing the alloy to 600 °C, interdiffusion of Al and Cr defines its structural and magnetic properties. Cr concentrates in the Fe-Cr-Co-rich A2 phase, which dominates the R2 islands and forms spheroidal magnetic precipitates in the R1 phase. A small amount of Cr can also be found in the Fe-Co-rich A2 shell. It is known from previous work that Cr aligns antiferromagnetically with Fe and Co [31] and reduces the magnetic moment. The new information on Cr accumulation in the A2 phase provides possibilities to vary the Cr concentration and annealing conditions to control the phase assemblies and magnetic properties. The application of stress and magnetic fields during annealing can also be used to control magnetic anisotropy.

The constituent phases in AlCo(Cr)FeNi HEAs have different magnetic properties, which form a rich and complex magnetic structure. Based on the measurements in this work, schematic representations of the deduced saturation magnetic induction and magnetic states are presented in Fig. 8. There are three primary contributions to the saturation magnetic induction from:

(i)     Fe-Cr-Co-rich A2 gen-2 and A2 gen-3 precipitates in region R1 and the A2 gen-1 matrix in region R2;

(ii)    Fe- Co-rich A2 shells between the A2 and B2 phases;

(iii)   the Al-Ni-rich B2 phase.

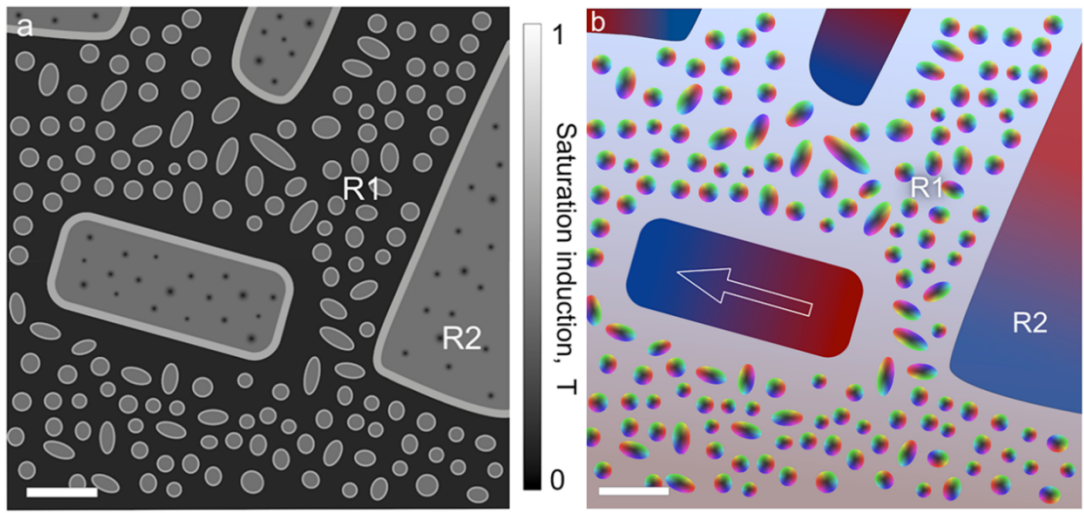

**Figure 8. Schematic diagrams of (a) magnetic induction and (b) magnetic states in the phase-separated AlCo(Cr)FeNi HEA.** The grayscale intensity in (a) corresponds to the deduced saturation magnetic induction in the different phases. Region 2 contains magnetic voids. The colors in (b) represent clockwise/ counterclockwise magnetic field rotation in magnetic vortices. a slowly-varying magnetic field in region 1 (R1), and large magnetic domains in region 2 (R2). The scale bar is 200 nm.

The magnetic state of each phase is distinctly different. The first two contributions are strongly linked, as they form core-shell structures with thin A2 shells around A2 spheres or islands. Off-axis EH measurements reveal that the A2 shell in region R2 (Fig. 7) has a higher magnetic induction than the A2 core (Fig.8a). Magnetic interactions, which can affect magnetization reversal, are expected between the two ferromagnetic A2 phases. The A2 spheres in region R1 support 3D magnetic vortex states in the B2 matrix. Based on the magnetic phase shift measurement and on the result of model-based iterative reconstruction of projected in-plane magnetization, the saturation magnetic induction in the A2 spheres, which have a core-shell structure, is estimated to be approximately 0.5 T (Fig. 5). For a 2.6-nm-thick shell around a core with a radius of 40 nm, the shell occupies almost 20% of the total volume. Therefore, the contribution of the thin Fe-Co-rich A2 shell to the total magnetization is significant. A magnetic contribution is also expected from the B2 phase, which contains more than 50% Fe, Ni and Co. In addition, the magnetic phase images (Fig. 6) provide evidence for a slowly-varying weak magnetic signal in the B2 matrix (Fig. 8b). The A2 spheres in region R1 support 3D magnetic vortex states in the B2 matrix (Fig. 8b), with a random distribution of clockwise and anticlockwise field rotations. Based on these observations, magnetization reversal and domain wall movement across the different magnetic components are expected to be highly complex processes.

In a multicomponent and multiphase alloy, the coercivity $H_c$ is expected to be sensitive to impurities, deformation, grain size and phase decomposition [32]. The AlCoFeNi HEA has a smaller value of $H_c$ than the AlCo(Cr)FeNi HEA. The magnetic structure in the AlCoFeNi HEA is characterized by large magnetic domains with small-angle magnetization variations. Magnetic domain walls move easily in the presence of an applied magnetic field. In contrast, in the AlCo(Cr)FeNi HEA, which has a multi-scale hierarchical B2+A2 decomposed structure, phase boundaries between the R1 and R2 regions can act as pinning sites for magnetic domain walls, thereby restricting their movement and resulting in an increase in coercivity $H_c$ compared to that in AlCoFeNi.

Our magnetic switching study shows that, in region R1, magnetic vortices in A2 gen-2 precipitates can change their rotation direction in the presence of external magnetic fields of 100 to 500 mT. The smaller the diameter of the A2 gen-2 precipitate, the lower is the magnetic field that is needed to change the magnetic vortex rotation direction, suggesting that the coercivity of such an AlCo(Cr)FeNi HEA can be tuned by changing the sizes and separations of the constituent A2 precipitates and islands.

In Al-Ni-Co (Alnico), a duplex nanoscale structure of two phases forms during thermal annealing in the presence of an external magnetic field. This leads to anisotropic growth of a periodic Fe-Co hard magnetic phase in an Al-Ni-rich matrix, resulting in shape anisotropy and enhanced coercivity. It is therefore of interest for future studies of AlCo(Cr)FeNi magnetic HEAs to determine the effect of external magnetic field and/or stress annealing, kinetics and other external stimuli to control magnetic anisotropy and magnetic properties, as already successfully demonstrated for Alnico. For instance, soft magnetic HEAs, in which moderate coercivity is tunable by magnetic nanostructures, are yet to be explored for high-frequency device applications.

**Conclusions**

The magnetic structure of AlCo$_x$Cr$_{1-x}$FeNi ($x$ = 0.5 and 1) heat-treated HEAs has been investigated with unprecedented spatial resolution in combined sub-Ångstrom structural and 3D chemical compositional studies. In AlCoFeNi, which contains nm-sized A2 Fe-Co-rich precipitates in a B2 matrix, the magnetic structure is characterised by large magnetic domains and small-angle magnetization variations. In contrast, the substitution of Co by Cr in AlCo(Cr)FeNi results in the formation of two regions: (i) a ferromagnetic A2 phase in a weakly-magnetic B2 matrix and (ii) B2 precipitates in a magnetic A2 matrix.

In the first region, the A2 precipitates are approximately spherical and, surprisingly, contain individual magnetic vortices. In the second phase, the B2 precipitates disrupt otherwise-continuous magnetic domains in the A2 matrix. The presence of an Fe-Co-rich A2 shell between the B2 and A2 phases provides an additional contribution to the overall magnetization. The saturation magnetization of the AlCo(Cr)FeNi HEA is dominated by the Fe-Cr-Co-rich A2 phases in both regions, as well as by the Fe-Co-rich A2 shells, whereas the B2 matrix phase provides a minor contribution. Its value is decreased by the substitution of Co by Cr as a result of the antiferromagnetic ordering nature of Cr.

The increased coercivity of the AlCo(Cr)FeNi HEA, which remains in the soft ferromagnetic range, can be attributed to a complicated magnetization reversal process, which involves the reversal of magnetic vortices in a weakly-ferromagnetic matrix. Our results provide direct local information about the intricate complexity of magnetic remanent states and reversal processes in multicomponent HEAs that contain coexisting magnetic phases and hierarchical structures that span multiple length scales.

**Methods**

**Specimen preparation.** AlCo$_x$Cr$_{1-x}$FeNi ($x$ = 0.5 and 1) bulk specimens were prepared by arc melting Al, Co, Cr, Fe and Ni pellets in an Ar atmosphere, followed by annealing at 600 °C for 15 h in an Ar atmosphere and quenching in water, as described elsewhere [20]. Bulk magnetometry measurements were performed in a vibrating sample magnetometer (VSM-Lakeshore 7404) using a maximum magnetic field of 1 T. Specimens for TEM and APT were prepared using focused Ga ion beam milling in FEI Helios Nanolab 400s and FEI Nova Nanolab 20 dual beam systems following a standard lift-out method. Electron-transparent (~100 nm) lamellae were attached to Cu Omniprobe support grids for TEM measurements.

**Atom probe tomography.** APT experiments were conducted in a LEAP 300X local electrode atom probe system (Cameca Instruments, Inc.). All atom probe experiments were conducted in electric field evaporation mode at a temperature of 60 K using an evaporation rate of 0.5% and a pulsing voltage of 20% of the steady-state applied voltage. Data analysis was performed using IVAS 3.6.2 software.

**Transmission electron microscopy.** HAADF STEM imaging, EDXS mapping and electron tomography were performed in an FEI Titan G2 80-200 electron microscope equipped with a high brightness field emission gun, a probe aberration corrector and an in-column Super-X EDXS system. HAADF STEM images were recorded on a Fischione detector using a beam convergence semi-angle of 24.7 mrad and an inner detector semi-angle of 69 mrad.

**Off-axis electron holography.** The same specimens that were used for microstructural characterization were used to study magnetic texture using Lorentz microscopy and off-axis EH. Off-axis electron holograms were recorded in magnetic-field-free conditions (*i.e.*, in Lorentz mode) in an image-aberration-corrected FEI Titan 80-300 electron microscope equipped with a high brightness field emission gun, an electron biprism, and a (Gatan K2 IS) direct electron counting detector [33] camera using a typical exposure time of 6 s. The biprism voltage was typically set to 100 V, resulting in an overlap interference width of 2.1 $\mu$m and a holographic interference fringe spacing of 2.76 nm with a contrast of 48% in vacuum. The objective lens of the microscope was used to apply out-of-plane magnetic fields to the specimen of between 0 and 1.5 T. The electrostatic and magnetic contributions to the phase shift were separated by turning the specimen over inside the electron microscope using a modified Fischione 2050 tomography specimen holder. Off-axis electron holograms were reconstructed numerically using a standard Fourier-transform-based method with sideband filtering using HoloWorks software in the Gatan microscopy suite, as well as using home-written scripts in the Semper image processing language [34]. Contour lines and colour maps were generated from recorded magnetic phase images to yield magnetic induction maps.

## Supplementary information

Supplementary Information accompanies this paper.

## Acknowledgments

Technical support from W. Pieper is gratefully acknowledged. This project has received funding from the European Research Council (ERC) under the European Union's Horizon 2020 research and innovation programme under grant agreements 856538 (3D MAGiC) and 823717 (ESTEEM3), as well as from the DARPA TEE program through grant MIPR# HR0011831554 and the Deutsche Forschungsgemeinschaft (DFG, German Research Foundation)-Project-ID 405553726-TRR 270. This work is also sup- ported by the AME Programmatic Fund of the Agency for Science, Technology and Research, Singapore under Grant No. A1898b0043.

## Author contributions

Q.L. and A.K. conceived and conducted the TEM experiments, S.V., R.B. and V.C. processed the alloys and conducted the APT and bulk measurements. Q.L, J.C., H.D. and D.S. analyzed the results. R.B., R.V.R. and R.E.D.B. provided research guidance. All authors contributed to the writing and reviewed the manuscript.

## Competing interests

The authors declare that they have no competing financial interests.

## Correspondence

Correspondence and requests for materials should be addressed to Q.L. (email: q.lan@fz-juelich.de) and A.K. (a.kovacs@fz-juelich.de).